\documentclass[pra,showpacs,superscriptaddress,byrevtex]{revtex4}

\usepackage{graphicx}
\usepackage{dcolumn}
\usepackage{amsmath}
\usepackage{epsfig}
\usepackage{bm}
\usepackage{amssymb}
\usepackage{array}
\usepackage{subfigure}

\newcommand{\sinc}{\ \mathsf{sinc}}
\newcommand{\ea}{\textit{et al }}

\begin{document}

\title{Decoherence in a double slit quantum eraser}
\author{F. A. Torres-Ruiz}
\author{G. Lima}
\author{A. Delgado}
	\affiliation{Center for Optics and Photonics, Universidad de Concepci\'on, Casilla 4016, Concepci\'on, Chile.}
    \affiliation{Departamento de Fisica, Universidad de Concepci\'on,  Casilla 160-C, Concepci\'on, Chile.}
\author{S. P\'adua}
	\affiliation{Departamento de F\'{\i}sica, Universidade Federal de Minas Gerais, Caixa Postal 702, Belo~Horizonte,~MG 30123-970,	Brazil.}
\author{C. Saavedra}\email{carlos.saavedra@cefop.udec.cl}
	\affiliation{Center for Optics and Photonics, Universidad de Concepci\'on, Casilla 4016, Concepci\'on, Chile.}
	\affiliation{Departamento de Fisica, Universidad de Concepci\'on,  Casilla 160-C, Concepci\'on, Chile.}


\date{\today}

\begin{abstract}
We study and experimentally implement a double slit quantum eraser in the presence of a controlled decoherence mechanism. A two-photon state, produced in a spontaneous parametric down-conversion process, is prepared in a maximally entangled polarization state. A birefringent double slit is illuminated by one of the down-converted photons, and it acts as a single-photon two-qubits controlled-\rm{NOT} gate that couples the polarization with the transversal momentum of these photons. The other photon, which acts as a which-path marker, is sent through a Mach-Zehnder-like interferometer. When the interferometer is partially unbalanced, it behaves as a controlled source of decoherence for polarization states of down-converted photons. We show the transition from wavelike to particle-like behavior of the signal photons crossing the double slit as a function of the decoherence parameter, which depends on the length path difference at the interferometer.
\end{abstract}

\maketitle

\section{Introduction}
\label{intro QE}

The complementary behavior of physical systems is at the heart of its quantum description. In brief, the principle of complementarity states that matter can exhibit physical properties that cannot be observed with certainty simultaneously \cite{Bohr1}, for instance,  particle-like and wavelike behaviors. The observation of particle-like and wavelike behavior in the well-known Young double slit experiment is a good example of the principle of complementarity. When the trajectory of the photons is recorded, the interference pattern disappears. On the other hand, when the trajectory followed by the photon is completely unknown, the interference pattern arises. For explaining this behavior, in a gedanken experiment, Einstein introduced the classical recoiling of the slit in his discussion with Bohr. In this thought experiment, the momentum transferred by the photons passing through a slit free to move transversally is recorded, and the interference should still be observed in the far field. However, Bohr refuted Einstein's experiment by using the uncertainty principle. For this, Bohr pointed out that for deducing through which slit the photon pass, the initial momentum of a free to move slit must be know within a momentum uncertainly that impose a position uncertainty that wash out the interference pattern. The notion of complementarity of some properties of physical systems remains valid \cite{Bohr2}. This type of complementarity is related to the preparation of the state preparation of the system.

The introduction of entanglement between the quantum system and a probe allows for the possibility of transferring system's information to a probe by choosing what measurement is being performed on the probe. In this context, the complementary behavior of the quantum system appears due to the entanglement between the interfering particle and the which-path marker particle \cite{Wooters1}. Scully and Dr\"{u}ll have studied the role of entanglement in a modified double slit experiment based on the photon emission of two atoms closely located \cite{Scully1}. They show that in some cases the loss of contrast in the interference pattern seems to be caused by the entanglement between the quantum system and the measurement device. The information of the markers can be deleted by properly selecting the measurement bases, i.e., the which-path information (WPI) can be erased from which-path markers (WPM) and the interference pattern is recovered. This erasure procedure is referred to as \emph{quantum eraser} \cite{Hillery,Scully2,Scully3}. In this sense, the state of the measurement device can be used as which-path marker. Then, when these states are orthogonal, the interference pattern is destroyed, even if they are not measured \cite{Steinberg1,Steinberg2}. The wave-particle duality may be quantified by relating interference fringe visibility and path knowledge as studied in Ref. \cite{Schwindt}. They have investigated this relation for states with different degree of purity. The WPI of a quantum system, of an entangled pair, can be marked or erased by its entangled twin even after the registration of the quantum \cite{Kim1}. This type of complementarity is associated to the measurement of noncompatible observables and is related to the Arthurs and Kelly uncertainty principle \cite{Arthurs}. A further discussion on the types of complementarity and the associated uncertainty product limit can be found in Refs. \cite{Bjork} and \cite{Trifonov}.

Several quantum eraser experiments have been done by employing two-photons states produced from the spontaneous parametric down-conversion (SPDC) process \cite{Schwindt,Herzog,Tsegaye,Kim1,Trifonov,Scarcelli,Walborn}. Here, we have used the photonic quantum eraser experiment introduced by Walborn \ea \cite{Walborn}. A related experiment has been reported by Pysher \ea \cite{Pysher}, where the birefringent element is a Mach-Zehnder interferometer in the path of the signal photon. Another related experiment has been reported by Gogo \ea \cite{Gogo} for comparing classical and quantum correlations by studying a quantum eraser based on a polarization interferometer.

In this article, we study how the purity of the entangled two-photon state in the double slit quantum eraser affects the wavelike and particle-like properties of the signal photons. In the experiment, down-converted photons are generated in a maximally entangled polarization state. Then the signal photon is sent through a birefringent double slit mask, which acts as a single-photon two-qubits controlled-\rm{NOT} gate that couples the polarization of the down-converted photons with the transversal momentum of the signal photons transmitted trough this aperture. This coupling allows the usage of the idler polarization states as a which-path marker. Thereby, selecting the orientation of a polarizer in the path of the idler photon of the entangled pair, the interference pattern can be erased or recovered. The control of the decoherence of the entangled pair is experimentally implemented by inserting an unbalanced Mach-Zehnder-like interferometer (MZI) along the propagation path of the idler photon, which acts on the WPI particle as an amplitude decay channel \cite{Nielsen}. The experimental setup is also suitable for studying the entanglement dynamics \cite{Walborn3}. Recently, we have implemented a erasure experiment in the context of state discrimination with partially entangled two-photon polarization states. In this case, the which-path marker states are not necessarily orthogonal. Hence, the visibility of the interference pattern as well as the WPI are constrained by the overlap between the which-path marker states \cite{Neves1}. By a probabilistic mapping of nonorthogonal states onto orthogonal ones, a complete WPI is obtained. However, when the mapping is onto collinear states interference with maximum visibility is recovered.

The remainder of this article has been organized as follows: In Sec. \ref{theory} we derive the expressions for describing the effects introduced by a controlled decoherence mechanism on the which-path marker states. In particular, we study how the decoherence process affects the visibility in the quantum eraser. In Sec. \ref{experiment} we describe the experimental setup and the obtained results. Finally, we conclude with a summary of the main results in Sec. \ref{summary}.

\section{Theoretical description}
\label{theory}

In this section we study the presence of a decoherence mechanism of a quantum erasure experiment in Ref. \cite{Walborn}. For sake of completeness and for an easy reading, we give a brief summary of the above mentioned reference. Thereafter, we study the use of a dynamically stabilized MZI, inserted along the propagation path of a single-photon field, as a source of controlled decoherence, where the degree of decoherence depends on the path length difference between the arms of the interferometer. Finally, we combine these results for studying the effect of a decoherence mechanism on a quantum erasure setup.

\begin{figure}[!tbh]
\centerline{\psfig{file=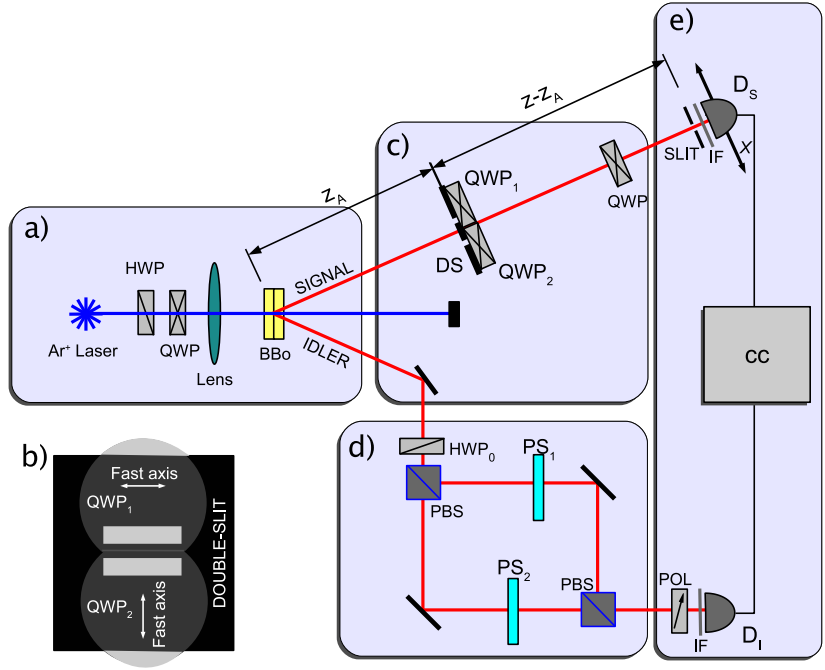,height=9cm}}
\caption{(Color online) Schematic representation of the experimental setup. a) The source of maximally entangled polarization state, Eq. (\ref{MEPS}), where the half-wave plate (HWP) and quarter-wave plate (QWP) allow for controlled generation of the state. A biconvex lens has been inserted to produce a large transversal coherence section for single-photons at the slit's plane located at $z_A$. (b) Alignment of the QWP's behind the double slit mask. (c) WPI section according the the protocol of Ref. \cite{Walborn}. The $QWP_{1}$ and $QWP_{2}$ are placed behind the double slit mask with the optical axis orthogonal between them. A final QWP is placed in the propagation path after the slits to change the polarization from $\left\{L,R\right\}$ to $\left\{H,V\right\}$. (d) The controlled decoherence mechanism. The idler photon pass through the actively stabilized MZI, where phase shifters ($PS_{1}$ and $PS_{2}$) provide the control over the path-length difference between the two arms of the interferometer. Finally, (e) detection system. The detectors are equipped with bandpass filters centered at $\lambda=702$ nm with $10$ nm at FWHM. Additionally, the signal detector is equipped with a single slit that enables the scan in the $x$ direction. In front of the idler detector, a linear polarizer determines the polarization basis for measurement.}
\label{QE:setup}
\end{figure}

\subsection{Quantum eraser protocol}
\label{qe_rev}

We assume that a two-photon state in a maximally entangled polarization state is generated in a SPDC process, as it is depicted in Fig. \ref{QE:setup}(a). The two-photon state is given by \cite{Kwiat}
\begin{eqnarray}
\left|\psi_{\rm in}\right\rangle=\frac{1}{\sqrt{2}}\left(\left|H_{i}V_{s}\right\rangle
+i \left|V_{i}H_{s}\right\rangle\right)\otimes\left|\psi_{\rm{spa}}\right\rangle,
\label{MEPS}
\end{eqnarray}
where $| H_j \rangle$ ($| V_j \rangle$) represents one photon in the $j$-th propagation mode with horizontal (vertical) polarization, with $j=s,i$ for signal and idler photons, respectively. The relative phase $\pi/2$ between the polarization components is generated due to the walk-off effect on the nonlinear crystals. The propagation modes of down-converted photons depend both on the pump beam spatial distribution and on the phase matching conditions \cite{Burnham,Keller,Souto}. These modes are correlated and $\left| \psi_{\rm{spa}} \right\rangle$ denotes the transversal spatial correlation of the two-photon state generated in SPDC.

The signal photon is passed through a vertically oriented double slit system. This double slit system can be interpreted as a filtering operation over the transversal momentum of signal photon. Neglecting a normalization constant, the filtered state, after the double slit, is given by
\begin{eqnarray}
\left|\psi_1\right\rangle \propto \left(\left|H_{i}V_{s}\right\rangle+i\left|V_{i}H_{s}\right\rangle\right)
\otimes\left|0_{i}\right\rangle\left(\left|+_{s}\right\rangle+\left|-_{s}\right\rangle\right),
\end{eqnarray}
where $\left|0_{i}\right\rangle$ is the spatial mode of idler photon defined by pinholes along its propagation path; $\left|\pm_s\right\rangle$ stands for the upper ($+$) and lower ($-$) slit's states. The $\left|\pm_s\right\rangle$ states are defined by
\begin{equation}
\left| \pm_s \right\rangle =\sqrt{\frac{\pi }{a}}\int_{-\infty
}^{\infty }dq_{s}e^{\mp idq_{s}}\mathrm{{sinc}\left(
q_{s}a\right) \left| 1_{q_{s}}\right\rangle } \label{eq:logical}.
\end{equation}
These states form an orthonormal basis for the Hilbert space of transmitted photons \cite{Neves2,Neves3}. Here, $a$ is the half width of the slits, $d$ is the separation between the center of the slits, $q_s$ is the signal photon transversal momentum, and the $\mathrm{sinc}$ function is the Fourier transform of the transmission function of the slits. A conditional operation can be performed over the polarization state of the signal photon, depending on the propagation path. For doing this quarter-wave plates (QWPs), with the optical axis perpendicularly oriented, after the upper and lower slits are inserted [see Fig. \ref{QE:setup}(c)]. The compound system, double slit and QWPs, acts as a single-photon birefringent double slit mask. The modified state is
\begin{eqnarray}
\left|\psi_2\right\rangle
\propto \left|H_i^{0}\right\rangle \!
	\left[\left|R_s^{+}\right\rangle-\left|L_s^{-}\right\rangle\right]
+ \left|V_i^{0}\right\rangle \!
	\left[\left|L_s^+\right\rangle+\left|R_s^-\right\rangle\right]
\label{LRB}
\end{eqnarray}
where we have used the following definitions $\left|R_s^{\pm}\right\rangle \equiv \left|R_s \right\rangle\otimes\left|\pm_s\right\rangle$; $\left|L_s^{\pm}\right\rangle \equiv \left|L_s \right\rangle\otimes\left|\pm_s\right\rangle$; $\left|H_i^{0}\right\rangle \equiv \left|H_i \right\rangle\otimes\left|0_i\right\rangle$ and $\left|V_i^{0}\right\rangle \equiv \left|V_i \right\rangle\otimes\left|0_i\right\rangle$, with $\left|R\right\rangle=(\left|H\right\rangle - i \left|V\right\rangle)/\sqrt{2}$ and $\left|L\right\rangle=(\left|H\right\rangle + i \left|V\right\rangle)/\sqrt{2}$. We can rewrite state (\ref{LRB}) as
\begin{eqnarray}
\left|\psi_2\right\rangle \!
\propto \!
 \left|L_i^0 \right\rangle\! \left[\left|P_s^{+}\right\rangle\!-\!i\left|P_s^{-}\right\rangle\right]
\!+\!i\left|R_i^0 \right\rangle\! \left[\left|M_s^{+}\right\rangle\!+\!i\left|M_S^{-}\right\rangle\right] \label{PMB}
\end{eqnarray}
with
$\left|P\right\rangle=(\left|H\right\rangle + \left|V\right\rangle)/\sqrt{2}$ and $\left|M\right\rangle=(\left|H\right\rangle - \left|V\right\rangle)/\sqrt{2}$.

From the above equations it is clear that by projecting idler photon onto the $\left|L_i\right\rangle$ ($\left|R_i\right\rangle$) state, interference fringes (antifringes) can be observed for the signal photon. In case of projecting onto $\left\{ | H \rangle, | V \rangle \right\}$ states there is no interference, i.e., it is possible to determine the path followed by the photons.

Because of the presence of the half-wave plate (HWP) oriented at $\theta=\pi/8$ in the path of the idler photon before the MZI, the
two-photon state is modified to
\begin{eqnarray}
\left|\psi_3\right\rangle\!\!\! &=& \!\!\!
 \left|M_i^{0}\right\rangle\!
	\left[\left|R_s^{+}\right\rangle\!-\!\left|L_s^{-}\right\rangle\right]
-\left|P_i^{0}\right\rangle\!
	\left[\left|L_s^+\right\rangle\!+\!\left|R_s^-\right\rangle\right]\label{PMB},\\
	&=&\!\!\!
  \left|L_i^0 \right\rangle \!
    \left[\left|P_s^{+}\right\rangle\!-\!i\left|P_s^{-}\right\rangle\right]
\!-\!\left|R_i^0 \right\rangle \!\left[\left|M_s^{+}\right\rangle\!+\!i\left|M_S^{-}\right\rangle\right]
\label{LRB2}
\end{eqnarray}

From Eqs. (\ref{PMB}) and (\ref{LRB2}), it can be seen that by projecting idler photon onto the $\left|L_i\right\rangle$ ($\left|R_i\right\rangle$) state, interference fringes (antifringes) are recovered. Now, in case of projecting onto $\left\{ | P_i \rangle, | M_i \rangle \right\}$
bases there is no interference.

We remark that, due to the mechanism used for implementing the which-path marking (birefringent double slit), the spatial interference depends on the measurement polarization basis of both the signal photon and idler photon; see for instance Eqs. (\ref{LRB}) to (\ref{LRB2}). However, as is pointed out by Kwiat \ea in Ref. \cite{Steinberg2}, it is pedagogically preferable that the which-path information be carried separately from the interfering particle. To satisfy this condition we must make a further projection of the signal polarization right after its transmission through the birefringent double slit \cite{Neves1}. Alternatively, we can choose to perform polarization measurement at idler basis in which the polarization of signal photon factorizes, as occurs in Eqs. (\ref{PMB}) and (\ref{LRB2}).

\subsection{Controlled decoherence mechanism}
\label{decoherencesp}

We assume a description for single-photons which take into account the frequency, propagation path and polarization degrees of freedom. In this case, we consider an initial propagation path $\left|0_i\right\rangle$ state and a superposition state of horizontal and vertical polarizations, denoted by $\left|H_i\right\rangle$ and $\left|V_i\right\rangle$, respectively. We assume here that generation of single-photon states is experimentally implemented by using two-photon polarization states in the degenerate case of spontaneous parametric down-conversion in a factorized state, namely we consider single-photon states and heralded detection. For this purpose, narrow-band Gaussian interference filters are inserted along the propagation path of down-converted photons, with a center frequency $\omega_c$ at the half of the frequency of the pump beam, i.e., $\omega_c=\omega_{p}/2$. At the experiment this is done by inserting the interference filters in front of the detectors. Under these assumptions, the initial state of idler photons is
\begin{eqnarray}
	\left|\varphi_{\rm in}\right\rangle
	&=&\int d\omega_i \phi(\omega_i)\left|1\omega_i\right\rangle\otimes\left(c_{h}
\left|H_i^0\right\rangle+c_{v}\left|V_i^0\right\rangle\right),
\end{eqnarray}
where $c_{h}$ and $c_{v}$ satisfy $|c_{h}|^{2}+|c_{v}|^{2}=1$ and $\phi(\omega)$ describes the spectral distribution of the photons, due to the presence of the interference filter, i.e., $\phi(\omega)\propto \exp[-(\omega-\omega_{p}/2)^2/2(\Delta\omega)^2]$.

Inserting a MZI along the propagation path of the idler photons, with polarizing beam splitter (PBS) at both the entrance and the output
of the interferometer, we implement conditional operations over the photons according to their polarizations. This is depicted in Fig. \ref{QE:setup}d. We have recently used a MZI for studying the unambiguous modification of nonorthogonal quantum states \cite{Torres}. The use of interferometers as controlled source of decoherence in entangled photons from SPDC has been reported \cite{Walborn2}. In this reference, Sagnac-like interferometers were used for implementing amplitude decay channels on single photons, for studying sudden death of entanglement induced by the environment.

In our case, we introduce different propagation lengths at the arms of the interferometer, which are denoted by $L_{a}$ and $L_{b}$. The postselected state at one of the outputs of the interferometer, neglecting a global phase, is given by
\begin{eqnarray}	
	\left|\varphi_{\rm out}\right\rangle =\int d\omega_i \phi(\omega_i)
    \left(c_{h}\left|H_i^0\right\rangle-c_{v}e^{i\frac{\omega_i \delta_I}{c}}
    \left|V_i^0\right\rangle\right)
	\left|1\omega_i\right\rangle\ \label{vp1}
\end{eqnarray}
where $\delta_I=L_{b}-L_{a}$ is the optical path difference between the propagation paths.

We assume that around the center frequency of the interference filters the response of the detectors is uniform. Hence, the state of the photon is obtained by tracing over the frequency domain. The reduced polarization density operator is given by
\begin{eqnarray}
	\!\!\!\!\!\!\!\!\rho_{i,\rm{pol}}&=& \rm{Tr}_{\omega}
\left( \left|\varphi_{\rm out}\right\rangle \left\langle \varphi_{\rm out}\right| \right) \nonumber \\
		&=& c_{h}^{2} \left|H_i^0\right\rangle\left\langle H_i^0\right|
		+c_{v}^{2} \left|V_i^0\right\rangle\left\langle V_i^0\right| \nonumber \\
		&-& \left(\gamma c_{h}c_{v}^*\left|H_i^0\right\rangle\left\langle V_i^0\right|+
        \gamma^{*} c_{h}^*c_{v} \left|V_i^0\right\rangle\left\langle H_i^0\right|\right),	
\end{eqnarray}
where $\gamma=\exp{\left(- \frac{\left(\Delta\omega\right)^{2}\delta_I^{2}}{2c^{2}} - i\frac{\omega_{c}\delta_I}{c}\right)}$. Hence, the MZI for photon polarization behaves as an amplitude decay channel. The  coherence length of the photons $l_{c}$ and the frequency bandwidth $\Delta \omega$ are related by $\Delta\omega=2\pi c/l_{c}=2 \pi c \Delta \lambda/\lambda^{2}$. With this, $\gamma$ can be expressed as a function of the dimensionless parameters $\varepsilon_{\lambda} $ and $\varepsilon_{I}$,
\begin{eqnarray}
	\gamma &=&\exp{\left(- 2 \pi^2 \varepsilon^2_{\lambda} \varepsilon^2_{I}
                    - 2 i \pi \varepsilon_{I} \right)}, \label{gamma}
\end{eqnarray}
where the dimensionless parameters are defined by $\varepsilon_{\lambda}= \Delta \lambda / \lambda$ and $\varepsilon_I=\delta_I / \lambda $.

The purity $\mathcal{P}=\rm{Tr}\left(\rho_{i,\rm{pol}}^{2}\right)$ of the polarization density operator is a function of the $\varepsilon$ parameters. The purity can be written as
\begin{eqnarray}
	\mathcal{P}= 1-2(c_{h}c_{v})^{2}\left(1- \left|\gamma\right|^{2} \right).
	\label{POP}
\end{eqnarray}
From this expression we can clearly see that in case of a balanced interferometer, $\left|\gamma\right|=1$, a pure state is recovered, i.e., $Tr\left(\rho^{2}\right)=1$. On the other hand, in case of an unbalanced interferometer, $\left|\gamma\right| < 1 $, a mixed state is obtained. The state is completely incoherent when $Tr\left(\rho^{2}\right)=0.5$. This occurs when $c_{h}=c_{v}=\frac{1}{\sqrt{2}}$ and $\left|\gamma\right|\rightarrow 0$. When we consider $\Delta\lambda=10$ nm and $\lambda=702.2$ nm, which is a typical configuration for the down-converted photons used in the SPDC experiments, then $\left|\gamma\right|\rightarrow 0$ when $\varepsilon_{I} \approx 30$.

\subsection{Field propagation}
\label{QE: propagacion de campo}

Here, we include a derivation of signal photon propagation from the double slit located at the $z_{A}$ plane up to the detector $z$ plane, as it is depicted in Fig. \ref{QE:setup}. At the latter plane a single slit located at a transversal $x$ position is placed in front of the detector $D_s$, the width of this slit is denoted by $2b$. We follow the derivation described in Refs. \cite{Neves2,Lima1,Lima2}.

We assume that the initial spatial state of the signal photon is given by
\begin{equation}\label{Spatial_sp}
\left|\Psi_s\right\rangle= W_{+}\left|+_s\right\rangle+W_{-}\left|-_s\right\rangle,
\end{equation}
with $\left|W_{+}\right|^{2}+\left|W_{-}\right|^{2}=1$ and the slits' states given by Eq. (\ref{eq:logical}). Hence, the field after the double slit at the $z$ plane is given by the spatial distribution, represented by the  function $F(q_{s})$ \cite{Lima2}. The state at this plane is then
\begin{eqnarray}
\!\!\!\!\!\left|\Psi_{z}\right\rangle&=&\sqrt{\frac{a}{\pi}} \int dq_{s} F(q_{s})
\left|1q_{s}\right\rangle, \label{Psiz} \\
\!\!\!\!\! F(q_{s}) &=&
	e^{-i q_{s}^{2}\alpha} \mathrm{sinc} \left(q_{s}a\right)
	\left( W_{+}e^{- iq_{s}d}+W_{-}e^{iq_{s}d}\right) \!, \label{FF}
\end{eqnarray}
where $\alpha=(z-z_A)/2k_s$ and $k_{s}$ is the transversal momentum of the signal photon. The first exponential in the above expression is the phase acquired due to the field propagation, the second one is due to the slits' positions, which are located at transversal positions $\pm d$, and the $\mathrm{sinc}$ function is the Fourier transform of the transmission function of the slits,
assumed to be a constant transmission of width $2a$. The slit in front of the detector at the $z$ plane can be described as a step function with constant transmission centered at transversal position $x$ with a width of $2b$. Thus, the transmitted signal state can be written as:

\begin{eqnarray}
\left|\Psi_{T}\right\rangle \!\!& \propto &\!\!
W_{+} \mathrm{sinc} \left(\frac{(x-d)a}{2\alpha}\right) e^{i\frac{\left(x- d\right)^{2}}{4\alpha}}\left|F_{-}(x)\right\rangle \nonumber\\
&&\!\!\! \!\!+
W_{-} \mathrm{sinc} \left(\frac{(x+d)a}{2\alpha}\right)e^{i\frac{\left(x+ d\right)^{2}}{4\alpha}}  \left|F_{+}(x)\right\rangle,
\label{Psit}
\end{eqnarray}
with
\begin{eqnarray}
\!\!\!\left|F_{\pm}(x)\right\rangle \!\!=\!\!\sqrt{\frac{b}{\pi}}\!\!\int\!\! dq_{s} e^{-iq_{s}x}\!\mathrm{sinc}\!\left(\frac{\left(x\pm d\right)b}{2\alpha}+q_{s}b\right) \!\left|1q_{s}\right\rangle.
\end{eqnarray}

The state of Eq. (\ref{Psit}) can be further approximated by considering the values of the experimental parameters $bd/2\alpha \simeq ad/2\alpha \simeq 10^{-3}$. Then, disregarding a global phase, the state at the detection plane is given by

\begin{eqnarray}
\!\!\left|\Psi_{T}\right\rangle\!\propto\! \left(
		W_{+}e^{-i \frac{\omega \delta_x}{2c}} \!+\! W_{-}e^{ i \frac{\omega \delta_x}{2c}}
		\right)\!\!\!\mathrm{sinc}\!\left(\frac{xa}{2\alpha}\right)\!\! \left|F(x)\right\rangle \!,\label{Psia}
\end{eqnarray}
where $ \delta_x =2 x d/(z-z_A)$ and
\begin{equation}
	\left|F(x)\right\rangle=\sqrt{\frac{b}{\pi}}\int dq_{s}
e^{-iq_{s}x}\mathrm{sinc}\left(\frac{xb}{2\alpha}+q_{s}b\right) \left|1q_{s}\right\rangle.
\end{equation}

\subsection{Decoherence effects in the double slit quantum eraser}
\label{main part}

Here, we use all the results of above subsections for studying the decoherence effects on the double slit quantum erasure experiment with two-photon states. The full two-photon state after crossing the birefringent double slit mask is given by:
\begin{eqnarray}
\!\!\!\!\!\!\!\left|\Psi\right\rangle \!\!&\propto& \!\! \int d\omega_i \Phi\!\left(\omega_i,\omega_{p}\!-\!\omega_i\right)\left|1_{\omega_{i}},1_{\omega_{p}-\omega_{i}}\right\rangle\! \times \nonumber \\
  & & \!\!\!\!\! \left(
            \left|H_i^{0}\right\rangle \left[\left|R_s^{+}\right\rangle-\left|L_s^{-}\right\rangle\right]
        + \left|V_i^{0}\right\rangle \left[\left|L_s^{+}\right\rangle+\left|R_s^{-}\right\rangle\right]
      \right). \label{Psid}
\end{eqnarray}

The spectral distribution of down-converted photons $\Phi(\omega,\omega_{p}-\omega)$ satisfies the phase matching conditions, i.e., in this case the energy conservation ($\omega_{s}=\omega-\omega_{i}$). Now, the idler photon is sent through a HWP oriented at
$\pi/8$ and afterwards through the MZI (as described in subsection \ref{decoherencesp}). Finally the idler photon arrives at the detector $D_{i}$. On the other hand, the signal photon propagates from the marked double slit as in secsection \ref{qe_rev} up to a single slit of width $2b$ centered at a transverse position $x$ in front of the detector $D_{s}$ at the $z$ plane. If we project spatial components onto
$\left|0_{i} \right\rangle \otimes \left|F_{s}(x)\right\rangle$, the final state is given by:
\begin{eqnarray}
	\left|\Psi_2\right\rangle \!\!\!&\propto&\!\!
	 \mathrm{sinc}\left(\frac{xa}{2\alpha}\right) \int d\omega
	\phi(\omega)^{2}\left|1_{\omega},1_{\omega_{p}-\omega}\right\rangle \times
	\nonumber \\
\!\!\!\!\!\!\!\!\!\!\!     && \left[
  \left(\left|R_{s}\right\rangle - e^{ \frac{i\omega_{s}\delta_{x}}{c}}\left|L_{s}\right\rangle\right) \left(\left|H_{i}\right\rangle+e^{\frac{i\omega_{i} \delta_I}{c}} \left|V_{i}\right\rangle\right) \right. \nonumber \\
    && \left.
\!\!\!\!\!\!\!\!\!\!\!    - \left(\left|L_{s}\right\rangle + e^{-\frac{i\omega_{s}\delta_{x}}{c}}\left|R_{s}\right\rangle \right) \left(\left|H_{i}\right\rangle
    -e^{\frac{i\omega_{i} \delta_I}{c}} \left|V_{i}\right\rangle\right)
    \right], \label{Psid2}
\end{eqnarray}
where we have assumed that the $\Phi\left(\omega_{i},\omega_{s}\right)$ spectral function slowly varies in a frequency window of width $2\Delta\omega$ around $\omega_p/2$. Identical Gaussian interference filters are placed in front of the detectors. The spectral transmission of these filters is described by the $\phi(\omega)$ distribution, and produces projective measurements over the whole state.
These filters are centered around half of the pump frequency with a width $\Delta\omega$. Besides, $\omega_{s}=\omega$, $\omega_{i}=\omega_{p}-\omega$; $\delta_I$ is the path difference along the MZI for the idler photon;  $\delta_{x}$ is the transverse path difference for propagation of the signal photon transmitted through the double slit up to the detector.

Until now, the state (\ref{Psid2}) is a pure state. Then the corresponding density operator is given by $\rho=\left|\Psi_2\right\rangle\left\langle \Psi_2\right|$.  If we trace over both the polarization of the signal photon and the frequency components, the reduced density operator for idler polarization is given by
\begin{eqnarray}
\rho_{\rm{pol}}  &=& \frac{1}{8}
\left(A_{-}\left\vert H_{i}\right\rangle \left\langle H_{i}\right\vert
+A_{+}\left\vert V_{i}\right\rangle \left\langle V_{i}\right\vert\right)\nonumber\\
&&-\frac{1}{8}\left(
B\left\vert H_{i}\right\rangle \left\langle V_{i}\right\vert
+B^{\ast}\left\vert V_{i}\right\rangle \left\langle H_{i}\right\vert\right).
\label{MDd}
\end{eqnarray}
where
\begin{eqnarray}
A_\pm \!\!\!&=&\!\!\! 2 \mathrm{sinc}^2 \!\left(\frac{xa}{2\alpha}\right)\! \\
B \!\!\! &=&\!\!\!
\mathrm{sinc}^2\!\left(\frac{xa}{2\alpha}\right) \left( e^{\xi_{+}}-e^{\xi_{-} }\right),
\end{eqnarray}
with $\xi_{\pm}\!=\!- 2\pi^2\varepsilon^{2}_{\lambda} \left(\varepsilon_{x}\pm \varepsilon_{I} \right)^{2}\pm 2 i \pi \left(\varepsilon_{x} \mp \varepsilon_{I}\right)$ and $\varepsilon_{x}=\delta_{x}/\lambda$. In this basis, we can see that coherence terms show an explicit dependence on the length difference between the arms of the MZI  $\delta_{I}$, while diagonal elements only depend on the $x$ transverse position at the $z$ plane.

The coincidence counts of down-converted photons both when projecting spatial components onto $\left|0_{i} \right\rangle \otimes \left|F_{s}(x)\right\rangle$ and when tracing polarization of signal photon and frequency components is given by:
\begin{eqnarray}
C_{\left| \theta_i \right\rangle } \propto \rm{Tr} (\rho_{\rm{pol}}  \left|\ \theta_i \right\rangle\left\langle \theta_i \right| ),
\end{eqnarray}
where $\left| \theta_i \right\rangle = \cos \theta_i \left| H_i \right\rangle + e^{i \varphi} \sin \theta_i \left| V_i \right\rangle$, namely, the projection measurement on the which-path marker (idler photon) has been performed onto $\left| \theta_i \right\rangle$ polarization state. Hence, $C_{\left| \theta_i \right\rangle }$ is proportional to a linear combination of the coefficients $A_{\pm}$ and $B$ of density operator in Eq. (\ref{MDd}). The linear combination depends on the parameters of state $\left| \theta_i \right\rangle$ as we will see in the following cases at the measurement procedure subsection.

\subsection{Measurement procedure}

In this section we study the dependence of the interference  of signal photons,  by performing projective polarization measurements on idler photons when the controlled decoherence mechanism is present.

\subsubsection{Measurement in the $\left\{\left|H\right\rangle,\left|V\right\rangle\right\}$
polarization basis.}

If we perform a measurement in the $\{\left|H\right\rangle,\left|V\right\rangle\}$ basis of the idler photon, according to Eq. (\ref{MDd})
only a diffraction pattern can be observed. This case corresponds to $\left| \theta_i \right\rangle$ with $\theta=\varphi=0$. For example, if a projective measurement on the $\left|H\right\rangle$ or $\left|V\right\rangle$ state of the idler photon is performed, then the spatial probability distribution along the $x$ direction at the $z$ plane is given by
\begin{eqnarray}
	C_{\left| H \right\rangle (\left| V \right\rangle)} & \propto& \mathrm{sinc}^2\left(\frac{xa}{2\alpha}\right).\label{QE: CHH}
\end{eqnarray}
In this basis, the coincidence count is independent of the optical path difference along the MZI.

\begin{figure}[t]
\centerline{
	\psfig{file=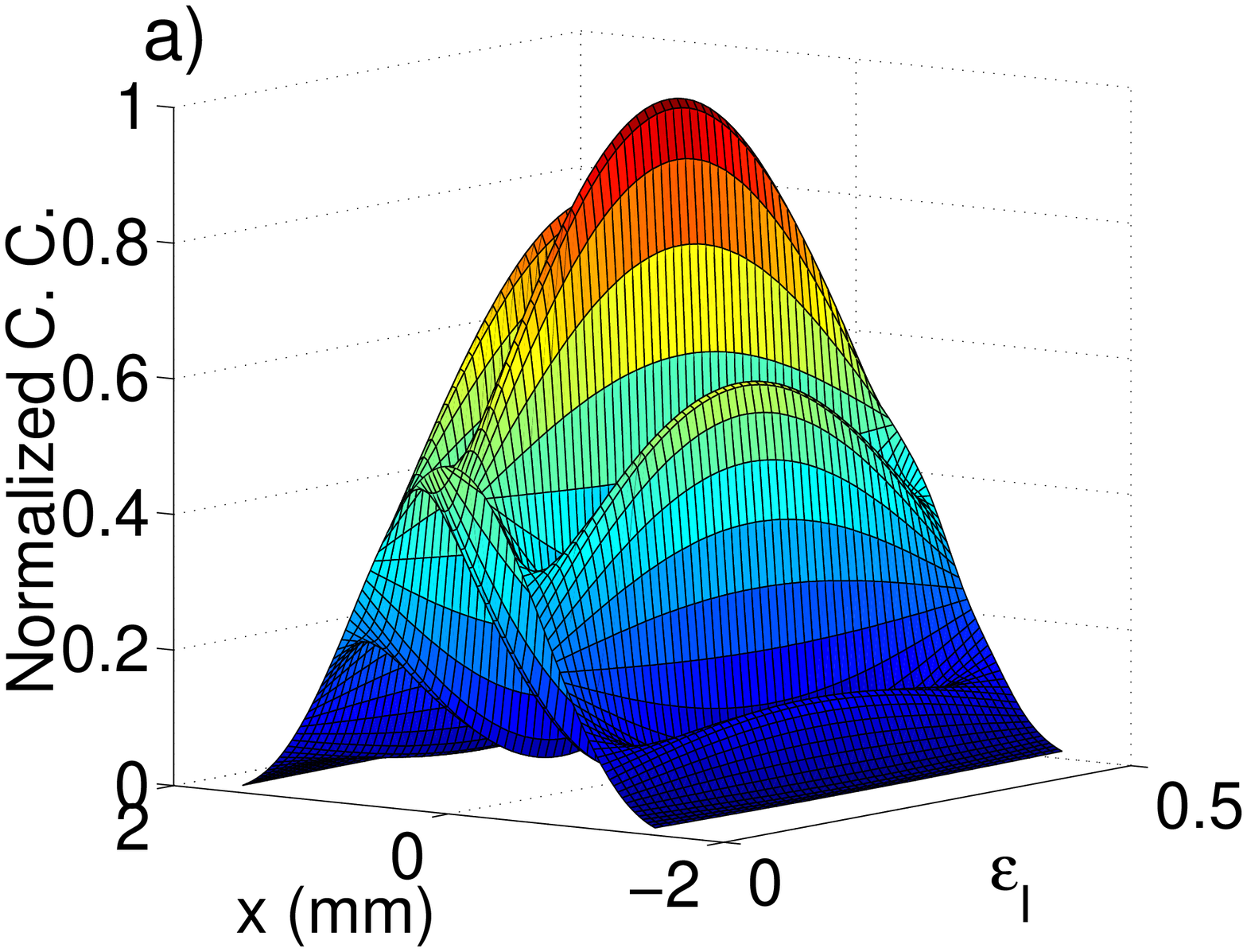, height=6.5cm, width=5cm ,angle=0}
	\psfig{file=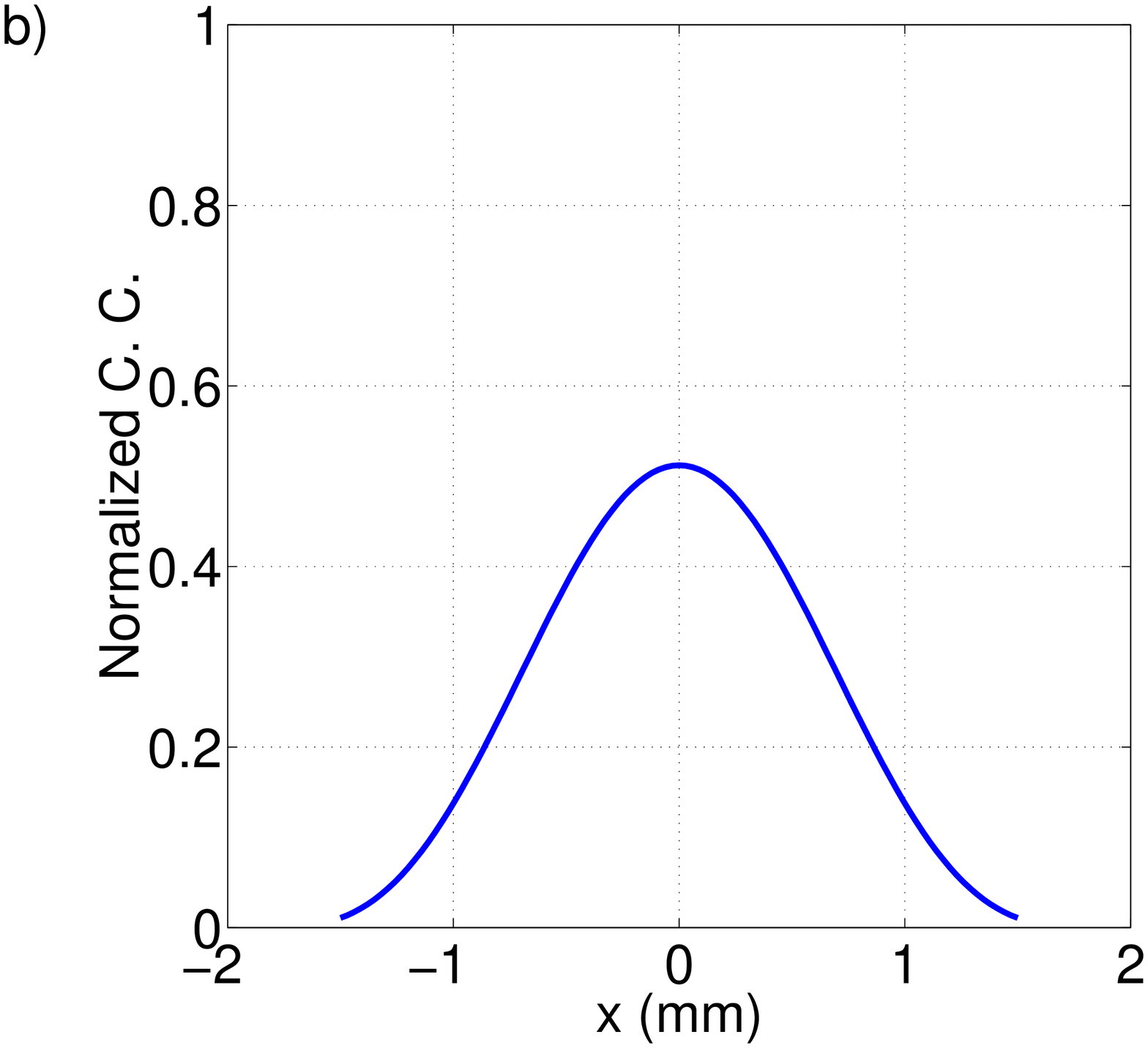, width=5cm ,angle=90}}
\centerline{
	\psfig{file=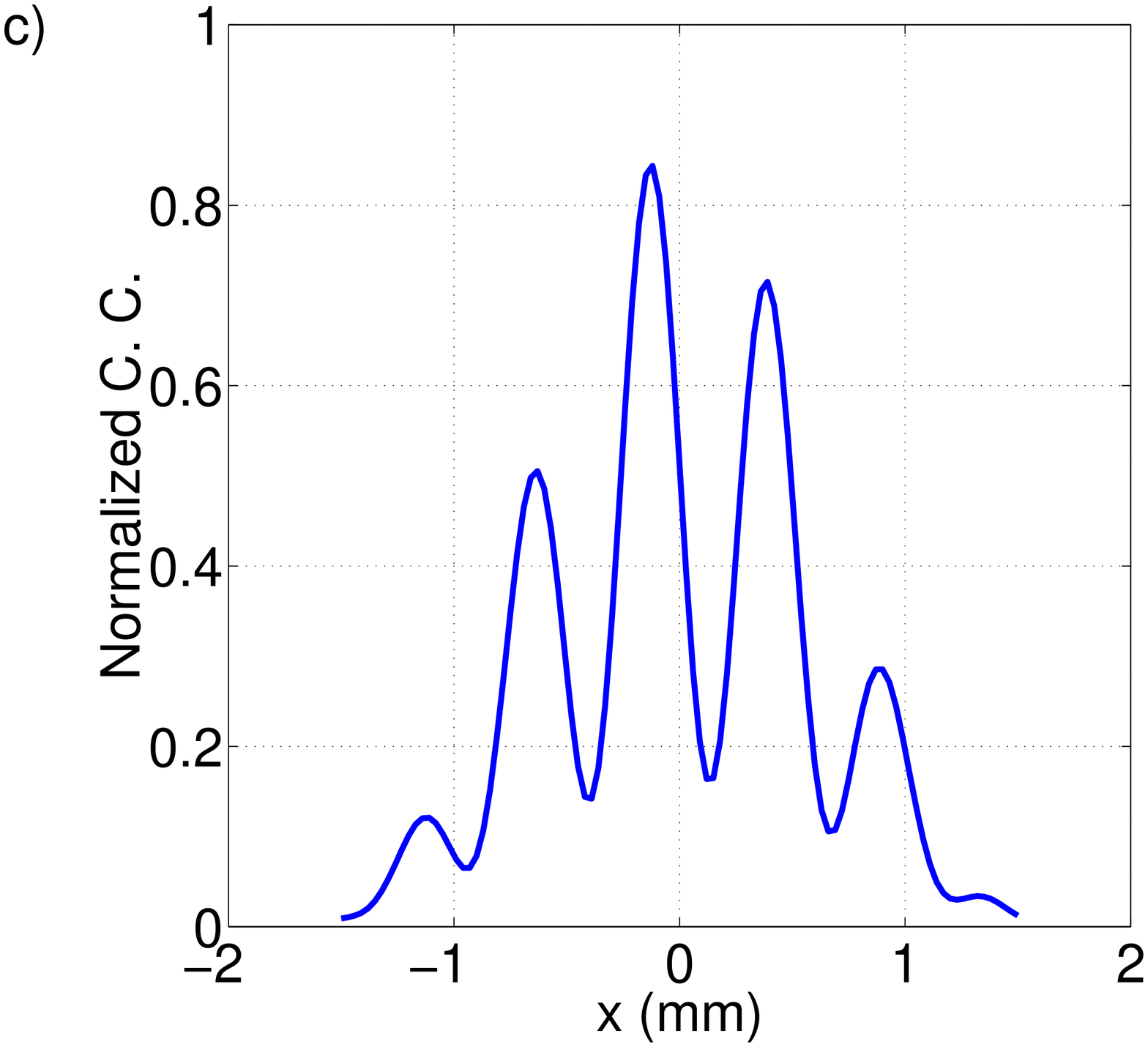, width=5cm ,angle=90}
	\psfig{file=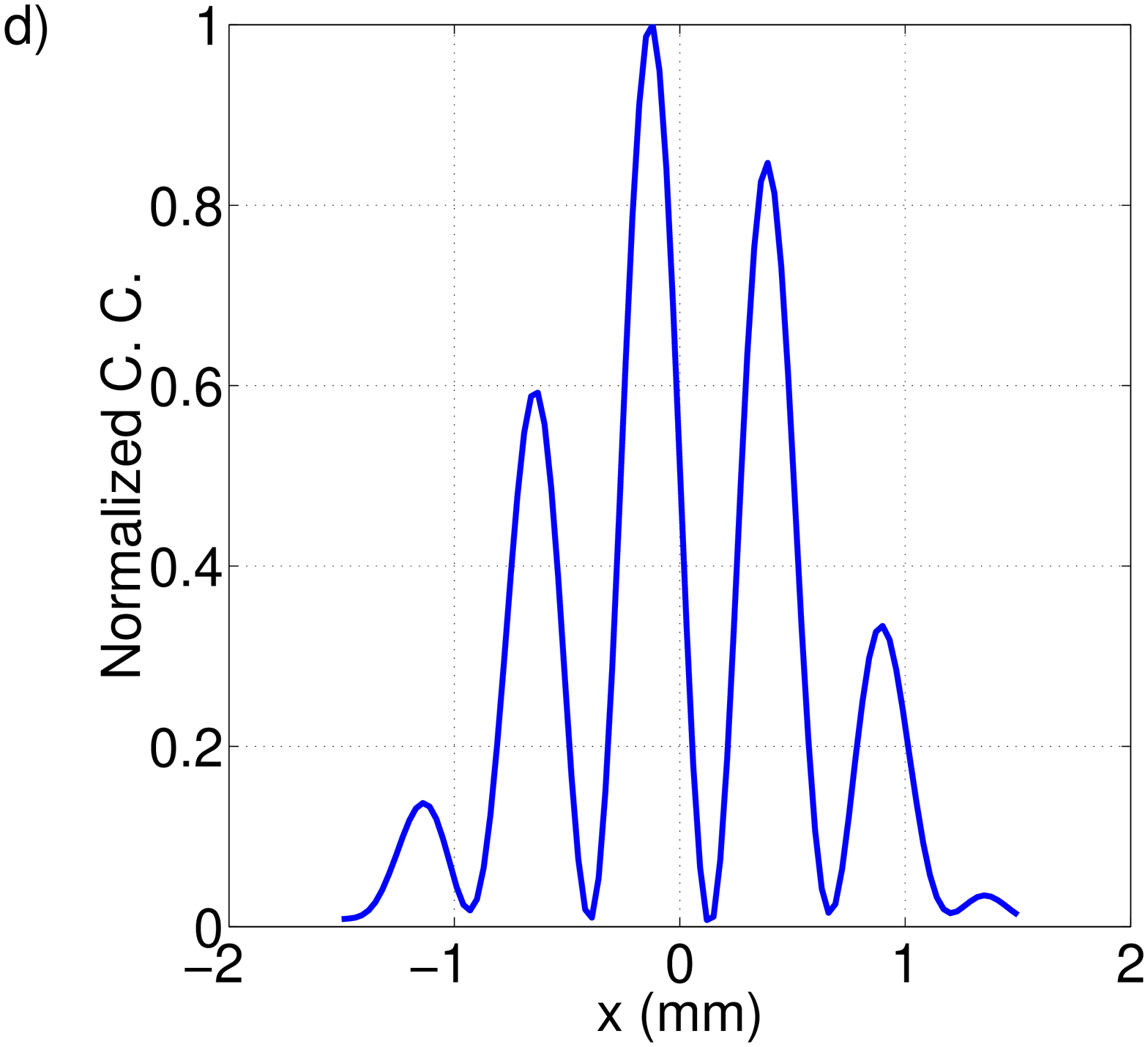, width=5cm ,angle=90}}
\caption{(Color online) Interference pattern of signal photon at $z$ plane, when idler photon is measured in $\left|P\right\rangle$, through normalized coincidence counts according to Eq. (\ref{Cp}). a) Three dimensional representation as a function of $x$ position at $z$ plane and $\varepsilon_{I}$. We can see that the maximum visibility of the interference pattern is obtained when $\varepsilon_{I}=1/4$, i.e., $V=1$. The other three curves show the interference patterns for different values of $\varepsilon_{I}$: (b) $0$, (c) $1/8$, and (d) $1/4$.}
\label{QE:deco-teo}
\end{figure}

\subsubsection{Measurement in the $\left\{P,M\right\}$ polarization basis.}
\label{M_pm}

Now, we study the case where the interference pattern is depending on the phase shift due to the length difference
between the arms of the MZI and the initial phase of the state, when the idler photon is projected onto the $\left| P \right\rangle$
and $\left| M \right\rangle$ polarization states. This case corresponds to $\left| \theta_i \right\rangle $ with $\theta=\pm \pi/4$ and $\varphi=0$. For this case we will write the reduced density operator in terms of the measurement
bases $\left\{	\left|P\right\rangle,\left|M\right\rangle\right\}$ and we will find a specific phase in the MZI to obtain the respective measurement basis that deletes the information.

Rewriting the density operator, Eq. (\ref{MDd}), in the $\left\{\left|P\right\rangle,\left|M\right\rangle\right\}$ basis, the  density matrix is given by
\begin{align}
\rho_{\rm pol}
&=
\frac{1}{8} \left(A^{(-)}-B-B^{*}+A^{(+)}\right)
\left|P_{i}\right\rangle\left\langle P_{i}\right| \nonumber\\
&+
\frac{1}{8} \left(A^{(-)}+B-B^{*}-A^{(+)}\right)
\left|P_{i}\right\rangle\left\langle M_{i}\right| \nonumber\\
&+
\frac{1}{8} \left(A^{(-)}-B+B^{*}-A^{(+)}\right)
\left|M_{i}\right\rangle\left\langle P_{i}\right| \nonumber\\
&+
\frac{1}{8} \left(A^{(-)}+B+B^{*}+A^{(+)}\right)
\left|M_{i}\right\rangle\left\langle M_{i}\right|.
\end{align}

Making a projective measurement onto the $\left| P \right \rangle$ polarization state, the probability of coincidence detection becomes
\begin{eqnarray}
C_{\left| P \right\rangle} & \propto & \mathrm{sinc}^2\left(\frac{xa}{2\alpha}\right)
\left(2- \left( \chi_{+} - \chi_{-} \right) \right), \label{Cp}
\end{eqnarray}
where
\begin{eqnarray}
\chi_{\pm} &=& e^{-2\pi \varepsilon^2_{\lambda} \left( \varepsilon_{x} \pm \varepsilon_{I} \right)^2 }
\cos\left(2\pi \left(\varepsilon_{x} \mp \varepsilon_{I} \right) \right). \label{chi}
\end{eqnarray}
In this case, the projection onto the $\left\vert M\right\rangle $ state has a $\pi$ phase
when compared to the $\left\vert P\right\rangle$ state.

From Eqs. (\ref{Cp}) and (\ref{chi}) we can make some important remarks. For example, we can see that the optical path difference along the MZI, $\delta_{I}$, appears both in the exponential and in the cosine function through the $\varepsilon_{I}$. Then, by varying this parameter, we can modify both the phase of the interference pattern and the contrast in the interference fringes simultaneously. This equation does not allow for a direct definition of a visibility of the interference pattern. Hence, we will analyze the results taking some particular cases.

\begin{figure}[!hb]
\centerline{\psfig{file=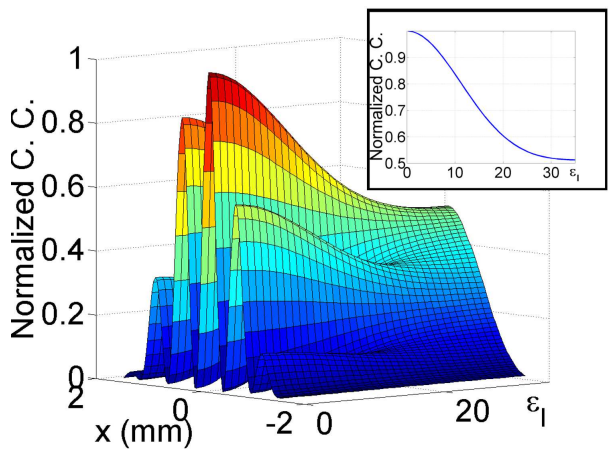, width=8cm ,angle=0}}
\caption{(Color online) Spatial distribution of the signal photon at the measurement  $z$ plane along the $x$ direction, as a function of $\varepsilon_{I}=n+\frac{1}{4}$, through normalized coincidence counts according to Eq. (\ref{cosenohiper2}). (Inset) The spatial distribution as a function of $\varepsilon_{I}$ for $x=0$, this corresponds to the maximum value of interference pattern . We observe that for $\varepsilon_{I}> 35$ the interference disappears and the result is a diffraction pattern.}
\label{QE:graficos 3d1}
\end{figure}

\begin{enumerate}
\item[a. (I)] \textbf{Unbalanced interferometer ($\varepsilon_{I}\gg\varepsilon_{x}$)}.

It can be seen in Eqs. (\ref{chi}) and (\ref{chi}) that for $\delta_{I}\gg \delta_{x}$ ($\varepsilon_{I}\gg\varepsilon_{x}$)  the exponential functions go to $0$ and the interference effects disappear. The resulting pattern consists in the diffraction of the  double slit with no interference. In this case the which-path marker state does not affect at all the double slit interference, so that only particle-like behavior can be observed.

\item[b. (II)] \textbf{Completely balanced interferometer ($\varepsilon_{I}=0$)}.

If we set $\delta_{I}=0$ ($\varepsilon_{I}=0$), then the interferometer is completely balanced and the exponential can be factorized. The cosine functions are canceled in this case. Thus, in this case no interference patterns are observed and we recover the diffraction of the double slit, as shown in Fig. \ref{QE:deco-teo}.(b).

\item[c. (III)] \textbf{Controlled displacements, $\varepsilon_{I}=\frac{n}{2}$.}

In the case when $\varepsilon_{I}=\frac{n}{2}$, with $n$ integer, the argument of the cosine functions is the same and we can factorize
this function. The resulting probability is given by
\begin{widetext}
\begin{align}
C_{\left| P \right\rangle} & \propto \mathrm{sinc}^2\left(\frac{xa}{2\alpha}\right)
\left(1+ (-1)^{n} e^{-2\pi^2 \varepsilon^{2}_{\lambda}\left(\varepsilon_{x}^{2}+n^{2}/4\right)}
\sinh\left(2n\pi^2 \varepsilon^{2}_{\lambda} \varepsilon_{x}\right)
\cos \left( 2 \pi \varepsilon_{x} \right) \right). \label{Cp2}
\end{align}
\end{widetext}

The experimental parameters for the argument of the $\sinh$ function is approximately $2n\pi^2 \varepsilon^{2}_{\lambda} \varepsilon_{x} \approx 0.0143 n$. We can see that, even for $n=10$, the contribution of this term is small and we can consider that the diffraction pattern is slightly modified but still is the double slit diffraction pattern.  Higher values of $n$ could not be relevant because the exponential function goes fast to zero as we shall see in the next case.

\item[d. (IV)] \textbf{Controlled displacements, $\varepsilon_{I}=n+\frac{1}{4}$}.

If we fix the value of $\varepsilon_{I}=n+\frac{1}{4}$, with $n$ integer, cosine functions can also be factorized, but we have
$\cos\left[2\pi \left((n+1/4) \pm \varepsilon_{x}\right)\right]=\mp \sin\left(2\pi\varepsilon_{x}\right)$. From this result we can see that expression (\ref{Cp2}) can be written as
\begin{widetext}
\begin{align}
C_{\left| P \right\rangle}& \propto
\sinc^2\left(\frac{xa}{2\alpha}\right)
\left(1- e^{-2\pi^2 \varepsilon^{2}_{\lambda}\left(\varepsilon_{x}^{2}+n^{2}\right)}
\sin \left( 2 \pi \varepsilon_{x} \right) \right), \label{cosenohiper2}
\end{align}
\end{widetext}
where we have approximated $\cosh\left[2(n+1/4)\pi^2 \varepsilon^{2}_{\lambda} \varepsilon_{x}\right]\simeq 1$ using the same values of the above studied case. In this way, we can see that the maximum of the interference visibility is obtained when the condition $\varepsilon_{I}=n+\frac{1}{4}$ holds as was shown in Fig. \ref{QE:deco-teo}.b. Also we can see that the phase of the fringes can be shifted by adjusting the value for $n$. Here, we remark that coincidence counts $C_{\left| P \right\rangle}$ under the condition $\varepsilon_{I}=n+\frac{1}{4}$ is equivalent to $C_{\left| L \right\rangle }$ with $\varepsilon_{I}=n$. The additional phase $\pi/2$ arising from the term $1/4$ in $\varepsilon_{I}$ when measuring in $\left\{P,M\right\}$ polarization basis can be seen as a measurement in $\left\{L,R\right\}$ polarization basis. Hence, this case corresponds to project polarization of idler photon onto $\left| \theta_i \right\rangle$ state with $\theta=\pi/4$ and $\varepsilon=\pi/2$.

\end{enumerate}

According to the results obtained in cases II and IV, the maximum (minimum) visibility of the interference pattern occurs when $\varepsilon_{I}=1/4$ ($\varepsilon_{I}=0$). Then we can vary the value of the dimensionless $\varepsilon_{I}=\delta/\lambda$ parameter from $0$ to $1/4$ for observing the modification of the visibility of the pattern. This is shown in Fig. \ref{QE:graficos 3d1}.

With these results we conclude the theoretical description of the experiment. In the next section we will discuss the implementation and the experimental results.

\begin{figure}[!bht]
\centerline{\psfig{file=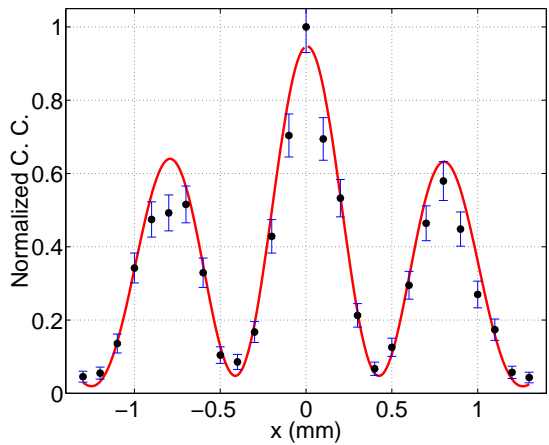, width=6cm ,angle=90}}
\caption{(Color online)
Normalized coincidence counts (C.C.) were recorded by scanning the signal detector in the $x$ direction at the $z$ plane.The signal photon passes through  the double slit array and the QWP's [as shown in Fig. \ref{QE:setup}(b)], while the idler photon propagates directly from the crystal to the detector. The maximum of coincidence was $205$ counts in $10$ s.}
\label{FRINC}
\end{figure}

\begin{figure}[!bht]
\centerline{
	\psfig{file=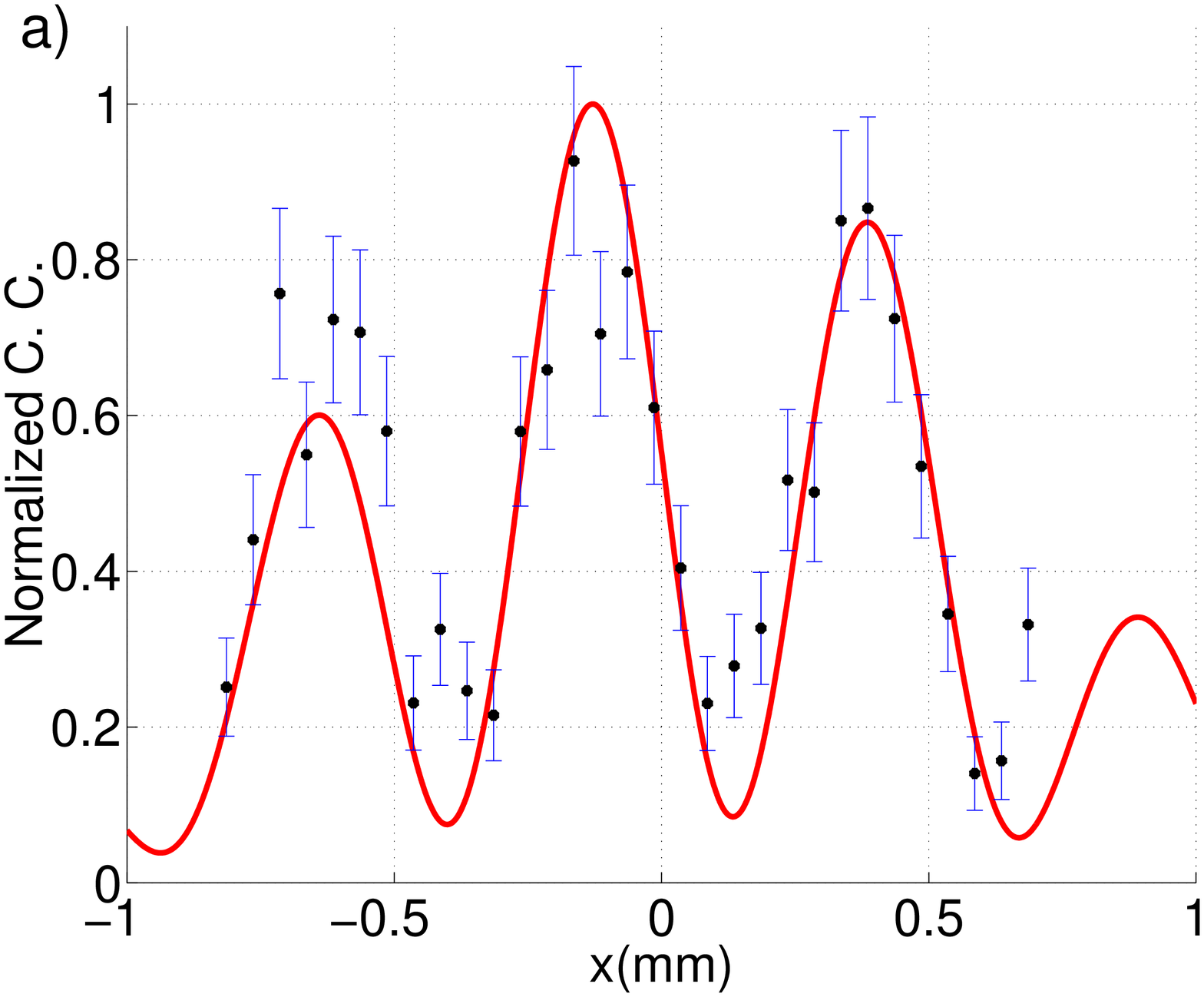, width=7cm ,angle=90}
	\psfig{file=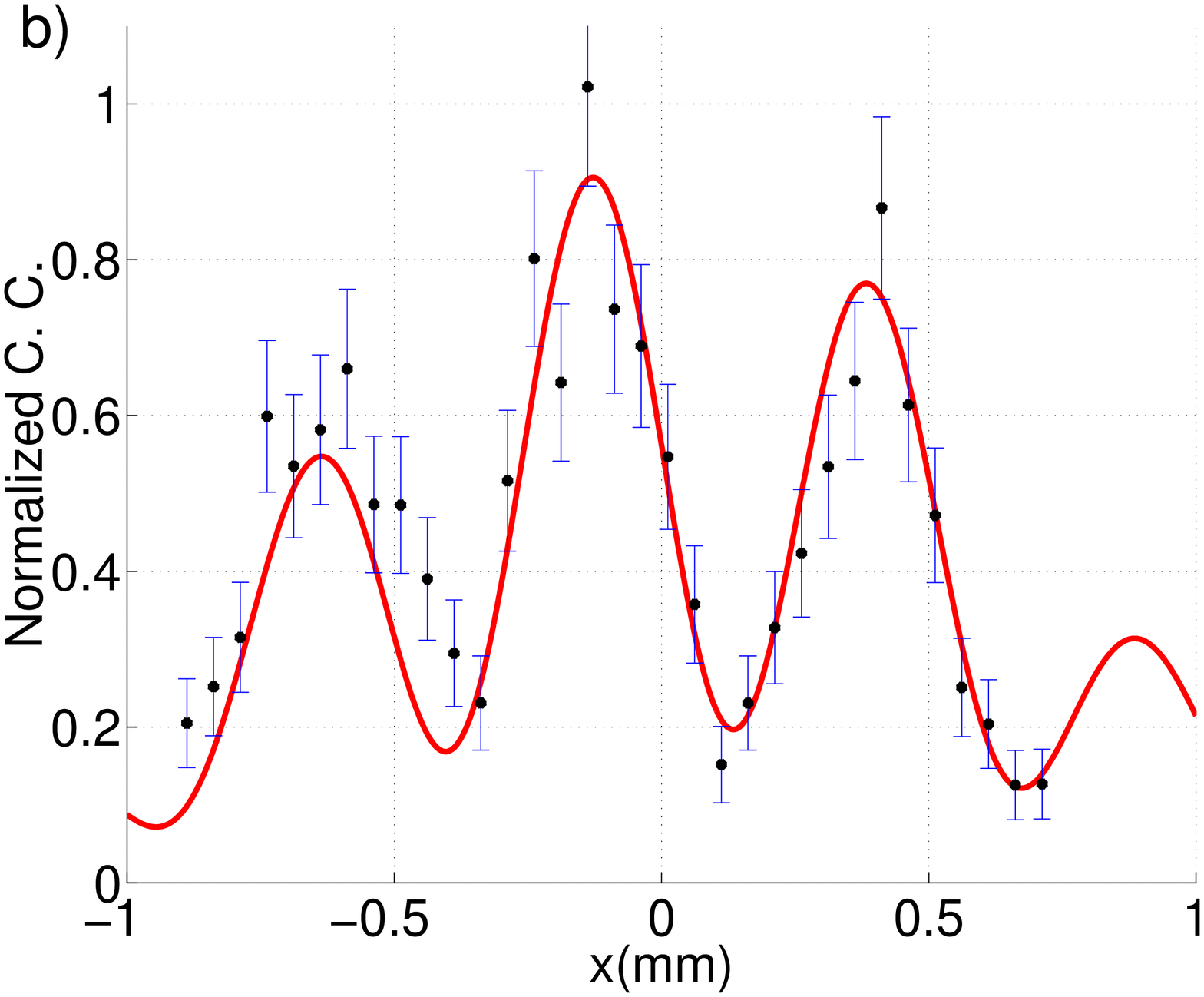, width=7cm ,angle=90}
	}
\centerline{
	\psfig{file=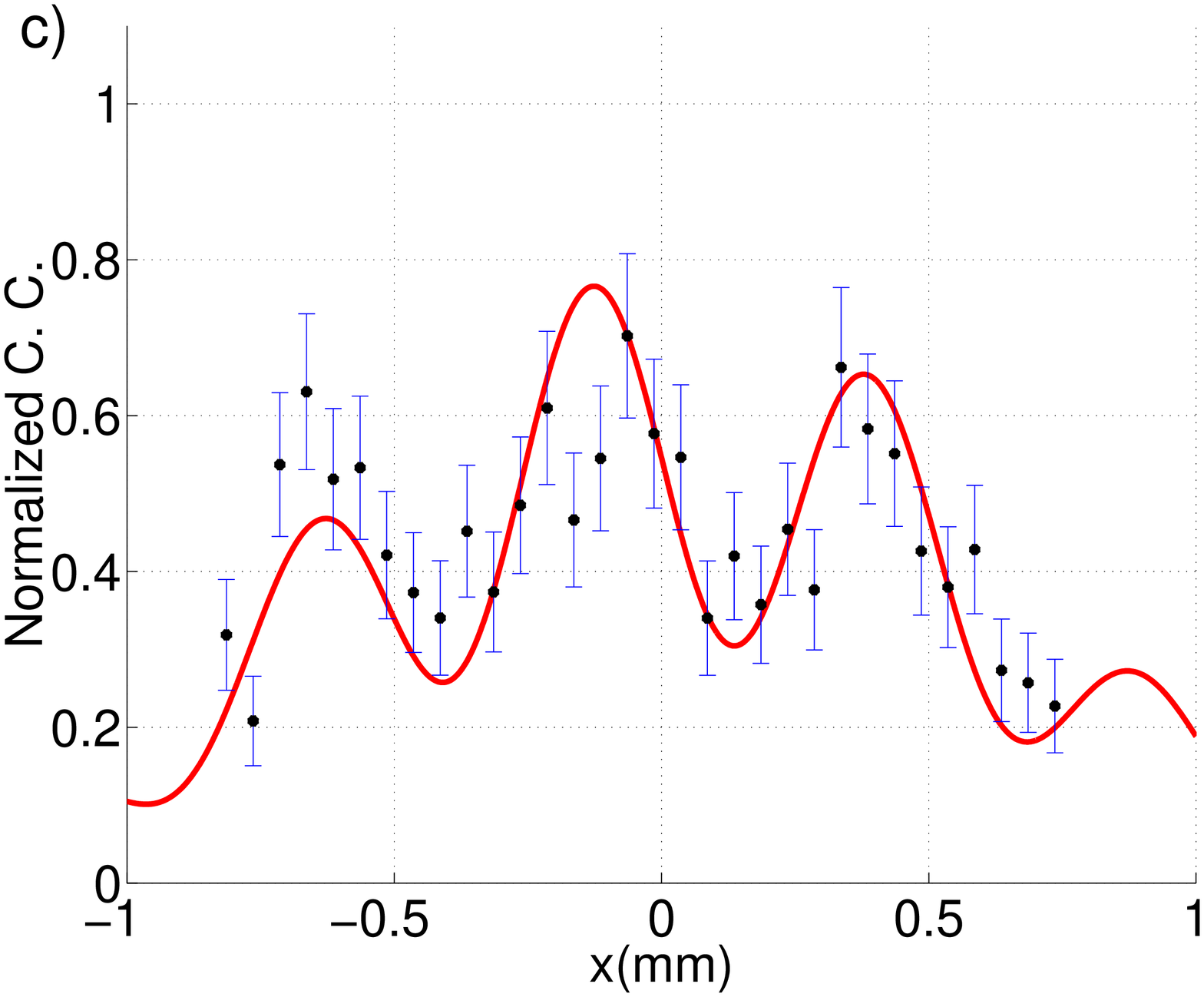, width=7cm ,angle=90}
	\psfig{file=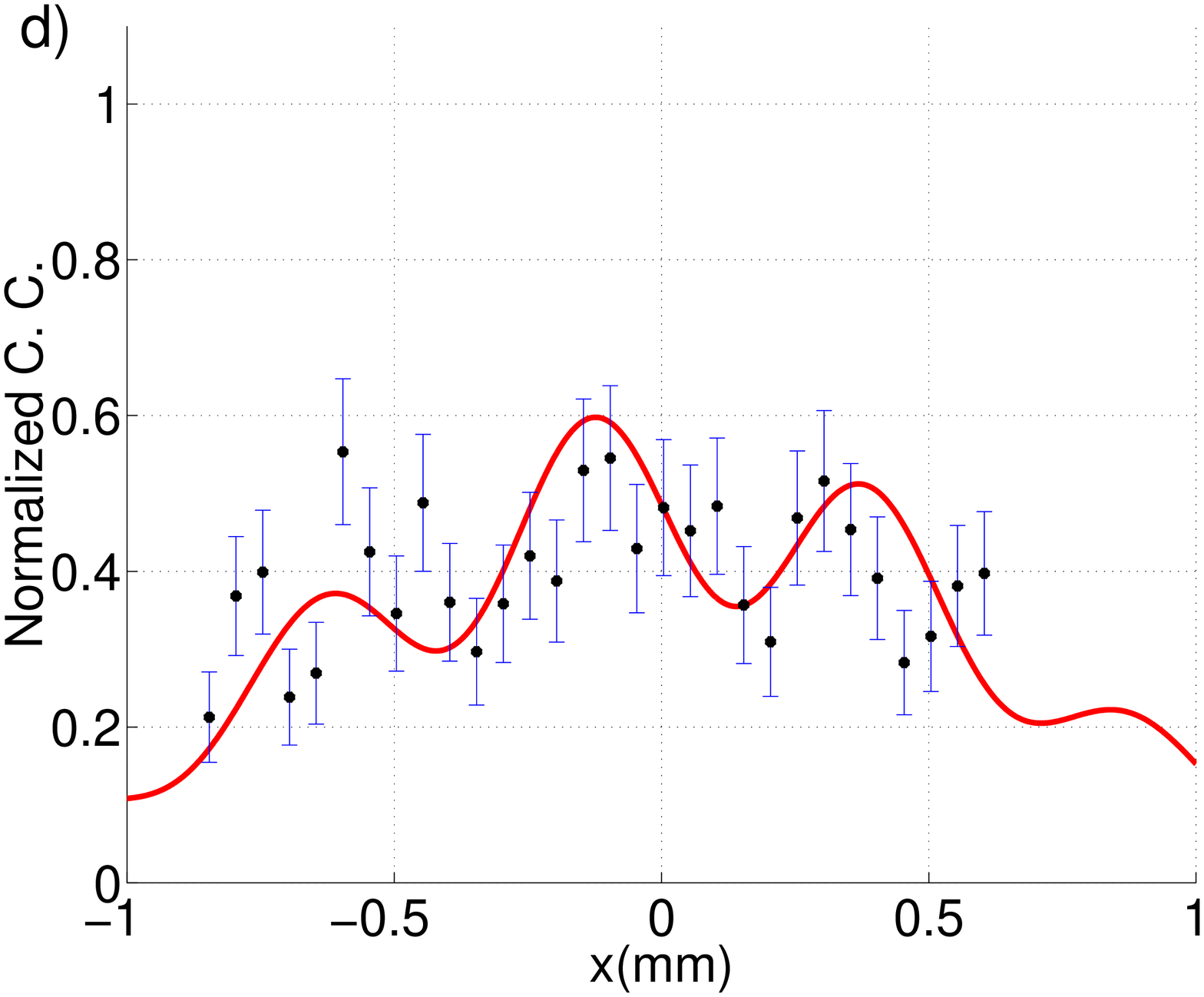, width=7cm ,angle=90}}	
\caption{\label{resultado interferencia} (Color online)
Normalized coincidence counts (C.C.) of the double slit interference patterns by scanning the signal detector in the $x$ direction at the $z $plane, while the idler photon propagates through the MZI, for different values of $\varepsilon_{I}=n+1/4$ parameter, with $n=$ (a) $7$; (b) $11$; (c) $15$ and (d) $19$. These values of $\varepsilon_{I}$ are belong to the case discussed in subsection \ref{theory}.2.IV. The dots (blue) correspond to the experimental results and the solid curve (red) is the theoretical fit given by Eq. (\ref{Cp}). The maximum of coincidence was $66$ counts in $10$ s.}
\end{figure}

\section{Experimental results}
\label{experiment}

In the experiment, we use a continuous wave (CW) Ar$^+$ ion laser working in a single-frequency mode at $150$ mW of power.
The polarization of the pump laser is controlled by using both a QWP and a HWP. The laser beam pumps two $1$-mm-wide
BBO crystals, which are cut for type I generation with their optical axes orthogonal. Thus, the crystals produce entangled pairs of photons in the state given by Eq. (\ref{MEPS}), where the phase and the amplitude of the state are controlled by properly setting the HWP and the QWP  shown in Fig. \ref{QE:setup}(a) \cite{Neves1}. Before the crystal, along the propagation path of the pump beam, a biconvex lens with a focal length of $30$ cm is inserted, with the crystals placed at the focal plane. This lens produces a large transversal coherence section for single-photons at the slit plane. The photons follow the propagation paths shown in Fig. \ref{QE:setup}(a).

The signal photon propagates from the BBO crystals to the double slit passing through a HWP oriented at $22.5^{o}$ that changes the polarization states from $\left|H\right\rangle$ ($\left|V\right\rangle$) to $\left|P\right\rangle$ ($\left|M\right\rangle$). This is required because the QWPs placed right after the slits have their optical axes parallel and perpendicular to the slits' direction. Therefore, the
QWPs axes and the polarizations form $45^{o}$ among them. The width of each slit is $a=80$ $\mu$m and the separation between them
is $d=250$ $\mu$m from center to center. The distance between the double slit plane and the detection plane was chosen as $z-z_{A}=0.2$ m. The detector placed at the propagation path of the signal photon is equipped with a single slit that is $50$ $\mu$m wide,  neutral filter, and  bandpass filter of $10$ nm at FWHM centered around $702$ nm. This detector scans the $x$ direction in order to record the interference pattern.

The idler photon propagates from the crystal to the MZI passing through a HWP ($HWP_{0}$ in Fig. \ref{QE:setup}) oriented at $22.5^{o}$, with the same change as in the signal photon. The input at the MZI is a polarizing beam splitter, which splits horizontal and vertical components of the idler photon. Inside the interferometer, we place a phase shifter that enables controlled modification of the path difference between the two arms of the interferometer. In practice, this is implemented by thin glass, $1.0\pm 0.01$ mm, plates inserted into each arm of the interferometer. The output of the interferometer is another PBS, which allows us to recombine the polarization components of the idler photon after a conditional phase shift. At the output of the interferometer we place a linear polarizer [POL in  Fig. \ref{QE:setup}(e)], which performs the projection measurement of the polarization of the idler photon. The quality of the MZI was tested by measuring the visibility of the single-photon interference pattern at the interferometer. For this purpose a product state was generated and the polarization of the input signal photon was rotated to $22.5^{o}$ and a linear polarizer rotated at an angle of $45^{o}$ was inserted in front of the detector $D_{s1}$. Interference pattern with visibility of $\mathcal{V}=0.94\pm 0.02$ was recorded by modifying the length of one of the interferometer's arm. This curve was obtained detecting in coincidence twin photons, where the signal photon propagates directly from the crystal to the detector while the idler photon passes through the MZI. The phase shifter was manually scanned. The length difference  at the two arms of the MZI was varied by using thin glass plates oriented at different angles. The variation of path length difference, $\delta$, is produced by the difference of rotation angles of these glass plates. These plates allow for a wide range of the $\delta$ parameter, more than $100$ times $\lambda$. These allows for setting the value of $\delta$ in a wide interval of values, from zero until $100$ times $\lambda$. To avoid fluctuations, the MZI was actively stabilized using a He-Ne laser beam propagating through the interferometer in the opposite direction to the idler photons. This beam produces interference fringes at the output, and a fast photon detector records the intensity of the fringes. This signal is used as a reference to stabilize the interferometer at any specific phase.

The detector $D_{i}$ is implemented with a bandpass filter and a neutral filter. Both detectors are connected to a coincidence detection circuit, where single and coincidence counts are stored, with a coincidence time window of $5$ ns. The calibration of the single-photon birefringent double slit mask  was performed using twin photons without the MZI in the idler path. This pattern is obtained scanning the transversal position in steps of $100$ $\mu$m, recording $10$ s for each measurement. The registered pattern is shown in Fig. \ref{FRINC}. A visibility of $\mathcal{V}=0.90\pm 0.02$ for this pattern is obtained. In case of a single double slit, the observed visibility was $\mathcal{V}=0.96\pm 0.02$. The transverse $x$ direction at the detection $z$ plane was scanned in steps of $30$ $\mu$m.

The controlled variation of the length difference between the arms of the MZI permits to study how the WPI at the measurement basis can be erased by varying the interferometer arms length difference. This study is implemented by varying the difference between the rotation
angles of the glass plates, in such a way that the $\varepsilon_{I} = n+1/4$. The double slit interference patterns are recorded by scanning the $x$ direction with detector $D_s$. The results of these scans are shown in Fig. \ref{resultado interferencia}. The blue dots correspond to the experimental data and the red curves are the best theoretical fit curves, which are given by Eq. (\ref{Cp}).
The curves are normalized by the highest value of coincidence counts (in this case $N_{cc}= 58$ counts in $30$ s) . We fit the best theoretical curve and we observe the highest value again considering the values obtained using the best fit. After this procedure, we normalize the curves using the highest value obtained. The phases for this fitted curves are [starting from (a) to (d)] $\varepsilon_{I} =\left(7+1/4\right), \left(11+1/4\right), \left(15+1/4\right), \left(19+1/4\right)$ and the angles of the glass plate are $\theta=0^{o}, 10^{o}, 16^{o}, 19^{o}$. Here, we remark that the measurement in the $\left\{P,M\right\}$ basis with $\varepsilon_{I} =n+\frac{1}{4}$ is equivalent to a measurement in the $\left\{L,R\right\}$ basis with $\varepsilon_{I} =n$.

We can see that the experimental results match well with the theoretical predictions. Actually, we can observe the transitions between wavelike and particle-like behaviors by controlling the length difference between the arms of the interferometer. Namely, the visibility of the double slit interference pattern decreases when this length difference increases.

\section{Conclusions}
\label{summary}

In this article we have both theoretically and experimentally studied the effects of a decoherence mechanism in the wave and particle-like behaviors in a double slit quantum erasure photonic setup. In particular, we have studied the effect of the decoherence mechanism in the measurement basis which exhibits the wavelike behavior with maximum visibility.

The decoherence mechanism affects the coherence terms of the reduced density operator for both single-photon and two-photon states. This mechanism is based on a conditional modification of the polarization of the idler. The conditional operation acting on polarization is implemented by means of a MZI, where the length difference between the arms of the interferometer add a conditional phase. This allows for going from a coherent to a completely incoherent state, in a fully controlled manner. This mechanism is suitable for studies of controlled decoherence in entangled polarization systems of photons. Furthermore, the incorporation of another MZI in the case of two-photon states allows a wider control of the degree of coherence, opening new possibilities concerning the controlled study of decoherence in optical systems.

We have observed that, in the implementation of the double slit quantum eraser, the WPI is deleted by properly selecting the measurement polarization basis. Besides, we have observed that the visibility of the interference pattern in a quantum erasure experiment not only depends on the choice of the measurement basis, but also is a function of the purity of the which-path marker state. In our case, even in the region of partial coherence (partially mixed density operator), it is possible to find specific values of $\varepsilon_{I}$
for deleting the which-path-information.

\section*{Aknowledgement}

We thank C.H. Monken for both interesting discussions and for lending us the sanded quarter-wave plates used in the double slit. This work was supported by Grants Milenio ICM P06-067-F and FONDECyT N$^o$ 1061046 and N$^o$ 1080383. S. P\'adua acknowledges the support of CNPq, FAPEMIG and National Institute of Science and technology in quantum information, Brazil.

\end{document}